\documentclass[aps,prl,twocolumn,showpacs,floatfix,superscriptaddress, longbibliography]{revtex4-2}
\usepackage[]{graphicx}
\usepackage{epstopdf}
\usepackage{ulem} 

\usepackage{graphicx}
\usepackage{dcolumn}
\usepackage{bm}
\usepackage{graphicx}
\usepackage{xcolor}
\usepackage{amsmath}
\usepackage{amssymb}
\usepackage{siunitx}
\usepackage{lipsum} 
\usepackage{enumerate}
\begin{document}

\title{High resolution quantum enhanced phase imaging of cells}

\author{Alberto Paniate}
\affiliation{Quantum metrology and nano technologies division,  INRiM,  Strada delle Cacce 91, 10153 Torino, Italy}

\author{Giuseppe Ortolano}
\affiliation{Quantum metrology and nano technologies division,  INRiM,  Strada delle Cacce 91, 10135 Torino, Italy}
\affiliation{Dipartimento di Fisica e Astronomia, Università di Firenze,
Via G. Sansone 1, I-50019 Sesto Fiorentino (FI), Italy}
\affiliation{Istituto Nazionale di Fisica Nucleare, Sezione di Firenze,
Via G. Sansone 1, I-50019 Sesto Fiorentino (FI), Italy}


\author{{Sarika Soman}}
\affiliation{Imaging Physics Dept. Optics Research Group, Faculty of Applied Sciences, Delft University of Technology, Lorentzweg 1, 2628CJ Delft, The Netherlands}

\author{Marco Genovese}
\affiliation{Quantum metrology and nano technologies division,  INRiM,  Strada delle Cacce 91, 10153 Torino, Italy}

\author{Ivano Ruo-Berchera}
\email{i.ruoberchera@inrim.it}
\affiliation{Quantum metrology and nano technologies division,  INRiM,  Strada delle Cacce 91, 10153 Torino, Italy}

\begin{abstract} 	

Recovering both amplitude and phase information from a system is a fundamental goal of optical imaging. At the same time, it is crucial to operate at low photon doses to avoid altering the sample, particularly in biological applications. Quantum imaging provides a powerful route to extract more information per photon than classical techniques, which are ultimately limited by shot-noise. However, the trade-off between quantum noise reduction and spatial resolution has long been regarded as a major obstacle to the application of quantum techniques to small cellular and sub-cellular structures, where they could offer the greatest benefits. Here, we overcome this limitation by demonstrating sub-shot-noise quantitative phase imaging of biological cells based on the transport-of-intensity equation, enabling high-fidelity, label-free imaging of key cellular and sub-cellular features. We achieve high-resolution phase imaging limited only by the numerical aperture, while simultaneously obtaining a resolution-independent quantum advantage. Unlike other quantum imaging approaches, our method operates in a quasi-single-shot, wide-field configuration, retrieves both phase and amplitude information, and does not rely on interferometric measurements, making it intrinsically fast and stable. These results pave the way for the immediate application of sub-shot-noise imaging in biological microscopy.

\end{abstract}	

\maketitle
\section{Introduction}
In biological imaging, many specimens, such as live cells and tissues, are nearly transparent under conventional light microscopy, making high-contrast imaging challenging without artificial labeling. Quantitative phase imaging (QPI) addresses this limitation by measuring optical path length differences, providing intrinsic contrast based on sample thickness and refractive index variations~\cite{Park_2018}.

In general, the extraction of meaningful information from both the amplitude and phase of an object critically depends on two key figures of merit: resolution and sensitivity. Resolution determines the smallest distinguishable detail in an image and is influenced by factors such as the optical system, the wavelength of light, and detector properties. Sensitivity, by contrast, refers to the ability to detect weak signals and is crucial for imaging faint biological structures. In optical imaging, sensitivity is often limited by photon quantum noise (shot-noise) and becomes a bottleneck when increasing the illumination level may induce phototoxicity or interfere with the biological processes under investigation \cite{Taylor_2016, cole2014live}. Improving resolution is often detrimental to sensitivity and vice versa, whereas maintaining an optimal balance between the two remains highly desirable.

Quantum sensing~\cite{Pirandola_2018,Degen_2017,Petrini_2020,Lee_2021,Polino_2020} and imaging~\cite{Defienne_2024,Berchera_2019,Genovese_2016} offer a solution to these limitations, with applications spanning from fundamental physics~\cite{Aasi_2013,Pradyumna_2020} to biological measurements~\cite{Taylor_2014,casacio2021quantum,Petrini_2022}, target detection~\cite{Tan_2008,Lopaeva_2013,Zhang_2015,Torrome_2024}, optical memory readout~\cite{Ortolano_2021a,Ortolano_2021b}, and pattern recognition~\cite{Ortolano_2023a}. 

In quantum imaging, single-photon emitters have enabled super-resolution fluorescence microscopy~\cite{Gatto_2014,Schwartz_2012,Tenne_2019}. Entangled photon pairs have been used for label-free spatial resolution enhancement~\cite{Toninelli_2019,defienne2022pixel,Zhang_2024,d2017characterization,Unternahrer2018}, quantum lithography~\cite{D'Angelo2001}, external background rejection~\cite{Lopaeva_2013,Defienne_2019,gregory2020imaging,Zhang_2024}. Quantum interferometric sensing approaches includes holography~\cite{Topfer_2022,Black_2023,Defienne_2021}, imaging with undetected photons~\cite{Lemos_2014,Kviatkovsky_2020,Valles_2018}, S(1,1) nonlinear interferometry~\cite{Machado_2020} and Hong-Ou-Mandel depth profiling~\cite{Devaux_2020,ndagano2022quantum}. However, these approaches do not enhance sensitivity beyond the shot-noise limit of their classical counterparts.
Phase sensitivity beyond the standard quantum limit can be achieved by exploiting $N$-photon entangled NOON states, which exhibit a de Broglie wavelength of $\lambda/N$~\cite{giovannetti2011advances}. Despite proof-of-principle experiments~\cite{Ono_2013,Israel_2014,Camphausen_2021}, practical implementations remain challenging due to the difficulties in generating NOON states with $N > 2$ and their susceptibility to optical losses and decoherence.

A more practical approach is provided by sub-shot-noise quantum imaging (SSNQI)~\cite{Brida_2010,Samantaray_2017,Sabines_2019,RuoBerchera_2020}, which enables sensitivity beyond the classical limit in weak object transmission measurements~\cite{Monras_2007,Adesso_2009,Nair_2018,Losero_2018,Wang_2024}. SSNQI exploits non-classical intensity correlations between two spatially separated beams~\cite{Jakeman_1986}: one beam interacts with the object forming a noisy classical image, while the other serves as a quantum noise reference, which is subtracted from the object image. Recent advances have extended SSNQI to non-interferometric quantum-enhanced phase imaging (NIQPI)~\cite{ortolano2023quantum}, where quantitative phase information is retrieved from intensity patterns using the transport-of-intensity equation (TIE)~\cite{Teague_1983,Paganin_1998,Zuo_2020}. A similar non-interferometric approach was recently employed for biphoton state reconstruction~\cite{dehghan2024biphoton}.
The strength of SSNQI and NIQPI lies in their favorable operating conditions: they do not require phase stability as in interferometric techniques, operate rapidly with  quasi-single-shot first-order intensity measurements, and are completely scanning-free. 
However, earlier demonstrations ~\cite{Brida_2010,Samantaray_2017,Sabines_2019,ortolano2023quantum} were limited by a significant trade-off between noise reduction and the lateral spatial resolution, which remains constrained by the biphoton correlation width at the object plane.

In this work, we introduce a high-resolution NIQPI approach that overcomes the resolution–sensitivity trade-off, achieving the first demonstration of quantitative sub-shot-noise phase imaging of biological cells at high spatial resolution. We identify and address the key mechanism underlying this advance: a previously unexplored interplay between noise propagation in TIE-based phase reconstruction and the specific properties of quantum noise subtraction. We provide a rigorous characterization of the experimental setup in terms of phase reconstruction fidelity, spatial resolution, and quantum advantage, using a pre-characterized nanofabricated sample featuring both phase and amplitude structures. We demonstrate a quantum advantage of approximately 30\% in reconstruction fidelity at a fixed illumination level. More importantly, we establish the immediate applicability of this technique by imaging unstained sea urchin ova, revealing cellular and sub-cellular features with a quantum-enhanced signal-to-noise ratio. Overall, this work bridges the gap between proof-of-concept quantum metrology and practical biological imaging, opening a concrete pathway toward the development of deployable quantum imaging technologies.

\begin{figure*}[t]
	\centering
	\includegraphics[height=0.41\textwidth,width=\textwidth]{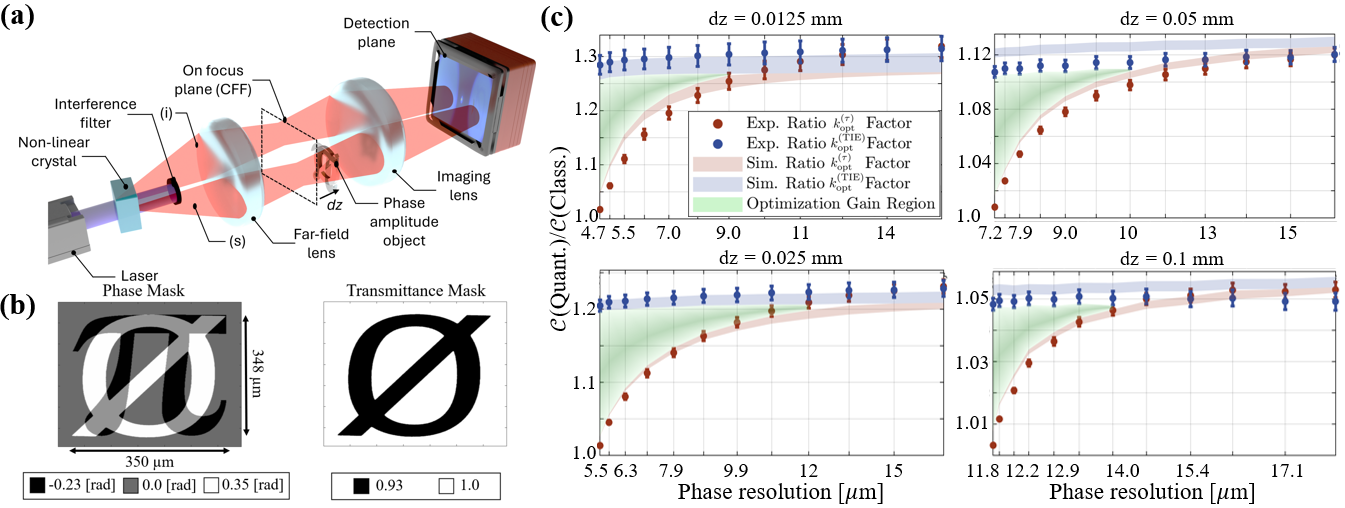} 
	\caption{\textit{High resolution quantum phase imaging technique} (\textbf{a}) Schematic of the non-interferometric quantum phase imaging setup~\cite{ortolano2023quantum}. Signal and idler beams generated by spontaneous parametric down conversion (SPDC) propagate through an \(f\)-\(f\) optical system, producing correlated intensity patterns. A phase-amplitude object, placed near the far field of the source, interacts only with the signal beam. Phase information is retrieved from intensity measurements placing the object at the focal and at the defocused planes (\(\pm dz\)). (\textbf{b}) Engineered test object consisting of distinct phase and transmittance masks. The phase mask consists of the superposition of a pure phase structure featuring a $\pi$-shaped symbol and a $\varnothing$-shaped symbol. The \(\varnothing\) symbol additionally features a transmittance of approximately 0.93. The phase-amplitude object has lateral dimensions of approximately \(348\,\mu\mathrm{m} \times 350\,\mu\mathrm{m}\), with thicknesses of about \(60\)~nm for the \(\pi\) structure and \(40\)~nm for the \(\varnothing\) symbol.  (\textbf{c}) The four panels show the experimental and simulated quantum advantage in phase estimation using the transmittance-optimized correction factor \(k^{(\tau)}_{\text{opt}}\)~\cite{ortolano2023quantum} and the newly proposed factor for TIE-based phase retrieval \(k^{(TIE)}_{\text{opt}}\) for four different defocusing distances \(dz\). The advantage is quantified as the ratio \(C(\text{Quant.})/C(\text{Clas.})\), where \(C(\text{Quant.})\) (\(C(\text{Clas.}))\) are the correlation coefficients between the quantum (classical) phase images and the reference one. The reported mean values and the associated uncertainty bars, corresponding to one standard uncertainty on the mean, are evaluated over an ensemble of approximately \(10^3\) images. The ratios, for the different distances \(dz\), are plotted as a function of the phase spatial resolution. Red and blue dots represent experimental data using \(k^{(\tau)}_{\text{opt}}\) and \(k^{(TIE)}_{\text{opt}}\), respectively, with shaded regions showing theoretical predictions. Green highlights indicate areas of optimization gain.}
\label{HRQPI_Scheme}
\end{figure*}

\subsection*{NIQPI}

The SSNQI and NIQPI protocols exploit the scheme depicted in Fig.~\ref{HRQPI_Scheme}\textbf{a}~\cite{ortolano2023quantum}. In a type-II nonlinear crystal pumped by a Gaussian coherent laser beam, spontaneous parametric down-conversion (SPDC) generates signal ($s$) and idler ($i$) beams, which become spatially separated in the crystal’s far-field (CFF), obtained at the focus of a lens in an $f$-$f$ configuration. Signal and idler photons are always generated in pairs along correlated and anti-symmetric directions leading to anti-correlated positions in the CFF. The spatial correlation in the CFF is characterized by a correlation length $l_{\mathrm{CFF}}$ ($l_{\mathrm{CFF}} \approx 5\,\mu\mathrm{m}$ in our setup \cite{Samantaray_2017}, approximately symmetric along both transverse directions, see M\&M for further details). The corresponding correlation area, $l_{\mathrm{CFF}}^2$, is approximately four orders of magnitude smaller than the overall spot size of each beam. In the low-gain regime of SPDC, the intensity patterns of the $s$- and $i$-beams are each characterized by shot-noise spatial fluctuations with a white spatial spectrum, i.e., the shot-noise is equally present at all spatial frequencies. Furthermore, the shot-noise fluctuations of the two beams are almost identically correlated (at  spatial areas larger than $l_{\text{CFF}}$) across the entire spot size, a feature impossible to reproduce with classical light sources.

The signal beam probes the object placed near the CFF plane. An imaging system projects the CFF onto a low-noise, high quantum-efficiency charge-coupled device (CCD) camera with magnification $M$, yielding photon counts $N_{s}(\mathbf{x},dz)$ in the pixel at the transverse coordinate $\mathbf{x}$, where $dz$ denotes the distance of the object from the CFF along the propagation axis. Hereinafter, we will indicate by $N'_{s}$, with ($'$), the intensity measured with the object and $N_{s}$ the one without the object.
Simultaneously, the idler beam is directed onto a separate region of the detector, producing a photon-count distribution $N_{i}(\mathbf{x})$, which is used to subtract the quantum-correlated shot-noise from the measured intensity in the signal arm. These intensity correlations can be directly applied in the SSNQI protocol. In particular, the transmission of an object placed in the CFF plane, $\tau(\mathbf{x})$, is estimated as~\cite{Moreau_2017, RuoBerchera_2020}:
\begin{equation}\label{tau_Q}
	\hat{\tau}_{Q} (\mathbf{x})= \frac{N'_s (\mathbf{x},0)- k^{(\tau)}_{\text{opt}} \delta N_i(\mathbf{x})} {\langle N_s (\mathbf{x},0)\rangle}.
\end{equation}
The quantity $\delta N_{i}(\mathbf{x}) = N_i(\mathbf{x}) - \langle N_i (\mathbf{x}) \rangle$ represents the fluctuation of photon counts in the correlated pixel of the $i$-arm and compensates for the fluctuation in the corresponding $s$-arm pixel. The factor $k^{(\tau)}_{\text{opt}}$ is chosen to minimize the variance $\langle \delta^2 \hat{\tau}_{Q} \rangle$, taking into account imperfect correlations, and will be discussed later. It is straightforward to show that this estimator is unbiased, with $\langle \hat{\tau}_{Q}(\mathbf{x}) \rangle = \tau(\mathbf{x})$.

NIQPI, presented in~\cite{ortolano2023quantum}, successfully applied SSNQI to image pure-phase samples using the TIE. While a pure-phase object produces no intensity contrast when precisely located in the focal plane of the imaging system, a slight displacement from this plane induces intensity variations due to propagation effects stemming from spatial phase gradients. The TIE method exploits these propagation-induced intensity changes to quantitatively retrieve the phase profile by solving the following equation:
\begin{equation}
\label{tie}
\left. -k\frac{\partial}{\partial z}I(\mathbf{x},z) \right|_{z=0} = \nabla_{\mathbf{x}} \cdot \left[ I(\mathbf{x},0) \nabla \phi(\mathbf{x},0) \right],
\end{equation}
which links the gradient of the transverse phase, denoted by $\phi(\mathbf{x},0)$, on the right-hand side to the axial derivative of the intensity on the left-hand side. The TIE formulation holds under the paraxial approximation and remains valid even for partially coherent illumination~\cite{Paganin_1998}.

In practice, the derivative in Eq.~(\ref{tie}) is approximated by a finite difference using two intensity measurements taken at planes symmetrically displaced by $\pm dz$:
\begin{equation}
\label{eq:approx}
\left. \frac{\partial}{\partial z} I(\mathbf{x}, z) \right|_{z=0} \approx \frac{I(\mathbf{x}, +dz) - I(\mathbf{x}, -dz)}{2dz}.
\end{equation}
In this sense, the technique is `quasi-single-shot' since only three  full-field intensity measurements performed at each plane (\(-dz, 0, dz\)) are required.
Ideally, choosing an infinitesimally small displacement ($dz \rightarrow 0$) would yield the most accurate approximation of the derivative. However, smaller values of $dz$ significantly reduce the strength of phase-induced intensity variations, making the signal indistinguishable from electronic and fundamental shot-noise. Conversely, larger displacements enhance noise resilience but reduce the accuracy of the finite-difference approximation, leading to out-of-focus phase reconstructions.
\begin{figure*}[t]
	\centering
	\includegraphics[height=0.37\textwidth,width=\textwidth]{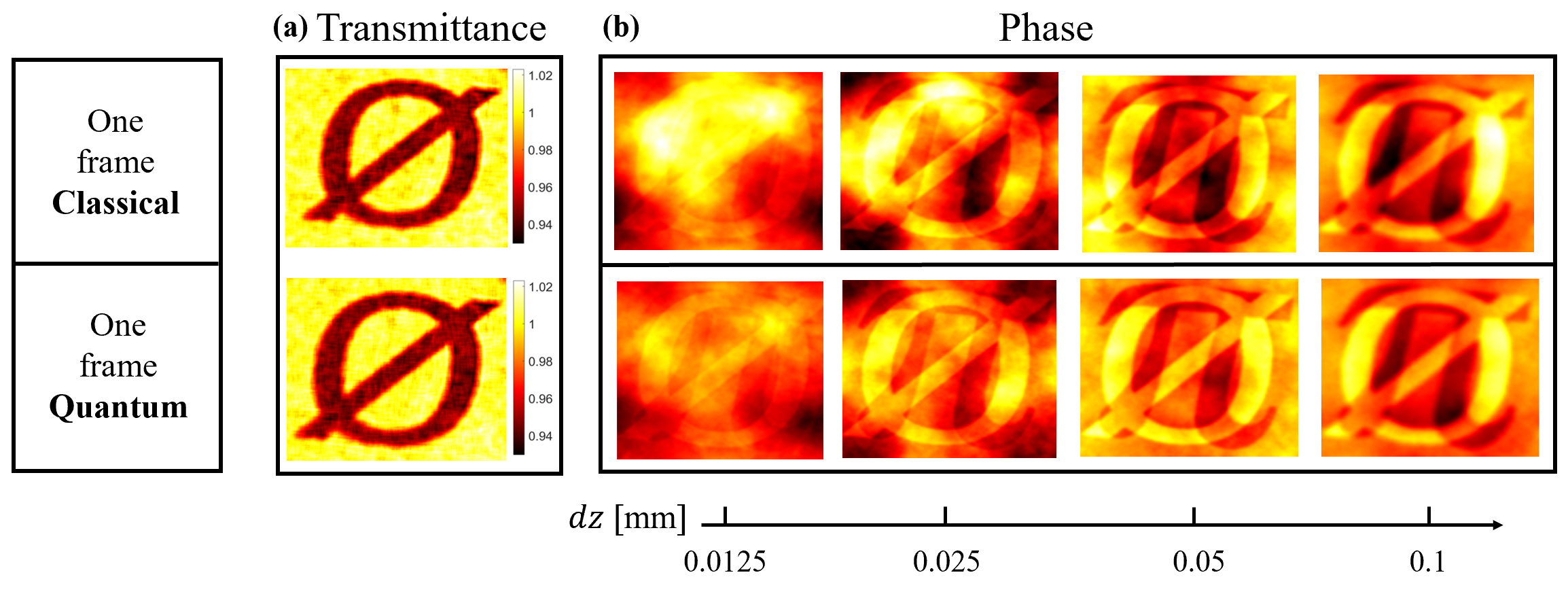} 
	\caption{\textit{Experimental reconstruction of transmittance and phase object} (\textbf{a}) Transmittance estimations for classical and quantum single-frame images, processed with a factor $\mathcal{D}=3.8$ obtained by averaging over 12 pixels. The object's area occupies $220 \times 220~\text{pix}^2$ with a mean number of photon per pixel of 600.  (\textbf{b}) Phase reconstruction at various defocus distances, $dz$, comparing classical and quantum single-frame approaches. The phase estimations at the smallest $dz$ are: classical $\pi = (-0.260 \pm 0.030$) rad, quantum $\pi = (-0.259 \pm 0.022)$ rad (theoretical \(\pi\) = $(-0.226 \pm 0.006)$ rad); classical \(\varnothing\) $= (0.37 \pm 0.02) $ rad, quantum \(\varnothing\) $= (0.36 \pm 0.02) $ rad (theoretical \(\varnothing\) = $(0.35 \pm 0.01)$ rad). Experimental uncertainties are obtained as mean errors calculated over \(10^3\) frames.} \label{fig:ClassicalQuantumLine}
\end{figure*}
Therefore, an optimal value of $dz$ must balance the accuracy of the derivative approximation against robustness to noise, and its choice critically depends on both the noise level and the specific phase profile of the sample. In this context, reducing shot-noise is essential for accurately approximating the intensity derivative via finite differences, thereby improving both the quantitative accuracy and qualitative clarity of the retrieved phase images. Details about the solution of Eq.~(\ref{tie}) are provided in the M\&M section.
 
In NIQPI, the quantum noise reduction is performed analogously to the SSNQI approach, replacing intensity terms in Eq. (\ref{tie}) with quantum-corrected intensities \cite{ortolano2023quantum}:
\begin{equation}\label{substitution}
    I(\mathbf{x},z)\mapsto N'_s (\mathbf{x},z)- k^{\text{(TIE)}}_{\text{opt}} \,\delta N_i(\mathbf{x},0),
\end{equation}
where the number of photon counts, \(N\), is considered in place of the intensities \(I\) since the two quantities are proportional, given a fixed detection time and area. The classical reference is obtained by the substitution \(I(\mathbf{x},z)\mapsto N'_s (\mathbf{x},z)\) without applying any noise subtraction, thus ensuring a fair comparison at equal photon dose. Note that, in the low-gain regime, each SPDC beam exhibits Poissonian spatial photon-number fluctuations, so this classical reconstruction represents the optimal classical benchmark. A description of the measured statistical properties is reported in the M\&M section.

While previous experiment focused exclusively on pure-phase samples \cite{ortolano2023quantum}, extending the approach to realistic biological applications introduces additional complexities. Biological specimens typically feature regions dominated by phase variations with negligible absorption, alongside areas where phase and absorption effects coexist. Accurately retrieving and distinguishing these two types of information is essential for a comprehensive characterization of both the structural and functional features of biological samples. 

To address this general scenario, we employed an engineered test object composed of distinct phase and transmittance masks (Fig.~\ref{HRQPI_Scheme}\textbf{b}). The phase mask consists of the superposition of a pure phase structure featuring a $\pi$-shaped symbol and a $\varnothing$-shaped symbol (more information is provided in the Supplementary Document). The respective phase shifts are $(\pi = -0.226 \pm 0.006)$~rad and $(\varnothing = 0.35 \pm 0.01)$~rad relative to the background @810~nm, the degenerate wavelength of the SPDC process. Additionally, the $\varnothing$-shaped symbol exhibits a transmittance of $(0.93 \pm 0.01)$ relative to the flat silica region.

The TIE method inherently addresses these challenges by decoupling phase-induced intensity variations from those arising due to the non-uniform transmittance~\cite{Zuo_2020}, thanks to the use of the intensity measurements acquired at symmetrically defocused planes, $I(\mathbf{x},\pm dz)$, as further described in the M\&M section.

\subsection*{Quantum Advantage vs Spatial Resolution}

The noise reduction factor (NRF), defined as \(
\text{NRF} = ~\frac{\langle \delta^2 (N_s - N_i) \rangle}{\langle N_s + N_i \rangle}, \)
is a well-established quantifier of non-classical correlations and shot-noise suppression, comparing the residual fluctuations in the difference of photon counts (numerator) to the shot-noise level expected for classical states (denominator)~\cite{Bondani_2007, brida2009measurement, jedrkiewicz2004detection, blanchet2008measurement, agafonov2010two}. For classical light, the NRF satisfies $\text{NRF} \geq~1$, while quantum-correlated states can reach values in the sub-shot-noise regime, $0 \leq \text{NRF} < 1$.

In the case of two correlated regions of a detector collecting SPDC photons, the NRF can be expressed as \(\text{NRF} = 1 - \eta_0 \eta_c(\mathcal{D}),\)
where $\eta_0$ is the single-channel detection efficiency, with the assumption of a balanced setup, $\eta_s = \eta_i = \eta_0$, and $\eta_c(\mathcal{D})$ represents the collection efficiency of correlated photon pairs~\cite{Samantaray_2017, RuoBerchera_2020}. The quantity $\eta_c(\mathcal{D})$ is an increasing function of the ratio \(\mathcal{D} = \frac{L_{\text{det}}}{M \cdot l_{\text{CFF}}},\)
with $L_{\text{det}}$ the linear size of the detector's integration area, $M$ the system magnification, and $l_{\text{CFF}}$ the correlation length in the crystal far-field. Further details about the estimation of \(\eta_0\) and \(\eta_c(\mathcal{D})\) are provided in the M\&M section. When \(\mathcal{D} \gg 1\), most of the correlated photon pairs are collected by two symmetric detection areas in the signal and idler beams, maximizing \(\eta_c(\mathcal{D})\approx 1\); conversely, for \(\mathcal{D} \lesssim 1\), only a fraction of the pairs are detected.

It is important to note that, while the spatial resolution $r_{\mathrm{CFF}}$ of the signal beam alone is inherently limited by the numerical aperture of the imaging system to $r_{\mathrm{CFF}} \approx 1.5\,\mu\mathrm{m}$ in our setup, noise subtraction in the intensity image is effective only on spatial scales larger than $l_{\mathrm{CFF}}$, that is for \(\mathcal{D} \gtrsim 1\). This introduces a fundamental trade-off between spatial resolution and quantum advantage in SSNQI for amplitude objects, based on the estimator in Eq.~(\ref{tau_Q}). This dependency has been extensively characterized in previous works~\cite{Samantaray_2017, Meda_2017, RuoBerchera_2020}. Under the assumptions of quasi-Poissonian photon number statistics in each arm and high object transmittance $\tau \approx 1$, the optimization factor $k^{(\tau)}_{\text{opt}}$ can be approximated as \(k^{(\tau)}_{\text{opt}}(\mathcal{D})\approx 1 - \text{NRF} = \eta_0 \eta_c(\mathcal{D}),\) indicating the dependence on the resolution parameter $\mathcal{D}$. In practice, $k^{(\tau)}_{\text{opt}}$ is calibrated through a preliminary characterization of the experimental NRF as a function of $\mathcal{D}$ (see M\&M).

At first glance, one might expect that the same resolution-related limitation would also affect the task of phase imaging via Eq.~(\ref{tie}), since the noise reduction in intensity measurements follows the same principle outlined in Eq.~(\ref{substitution}). However, the main goal of this work is to demonstrate that this is not the case: the quantum advantage in phase retrieval is largely independent of the spatial resolution of the reconstructed images.

To understand why this is the case, it is essential to consider how intensity noise propagates into the retrieved phase profile. By examining Eq. (\ref{tie}), under the assumption that the finite-difference term $I(\mathbf{x}, dz)-I(\mathbf{x}, -dz)$ is dominated by noise fluctuations $\sigma(\mathbf{x})$, and considering a nearly uniform illumination intensity $I(\mathbf{x},0)\approx I_0$, the equation simplifies to \cite{Paganin_2004}:
\begin{equation}
    -k\frac{\sigma(\mathbf{x})}{\sqrt{2}\,I_0\,dz} = \nabla^2 \phi_{\text{noise}}(\mathbf{x}).
\end{equation}
Considering the Fourier transform of both \(\sigma(\mathbf{x})\) and \(\phi_{\text{noise}}(\mathbf{x})\) yields:
\begin{equation}
\label{eq:noiseSuppres}
    \frac{k\,\tilde{\sigma}(\mathbf{q})}{4\pi^2\sqrt{2}\,I_0\,dz\,|\mathbf{q}|^2} = \tilde{\phi}_{\text{noise}}(\mathbf{q}),
\end{equation}
where $\tilde{\sigma}(\mathbf{q})$ and $\tilde{\phi}_{\text{noise}}(\mathbf{q})$ denote the Fourier transforms of the intensity noise and the resulting phase noise, respectively, and $\mathbf{q}$ represents the spatial frequency vector. In the case of shot-noise we can assume a flat spectrum, $\tilde{\sigma}(\mathbf{q})=\tilde{\sigma}$, for \(\mathbf{q} \neq 0\). The presence of the $|\mathbf{q}|^2$ term in the denominator highlights that the phase noise in the TIE reconstruction is dominated by low spatial-frequency components, i.e. by the slowly varying shot-noise fluctuations across the image. In contrast, high-frequency noise components, such as pixel-to-pixel shot-noise, are strongly suppressed by the intrinsic low-pass filtering nature of the TIE, and therefore have minimal impact on the retrieved phase. Incidentally, the remaining low-frequency components of the shot-noise are precisely those efficiently removed through quantum correlations, by the substitution in Eq.~(\ref{substitution}). 

As a direct consequence of the TIE noise propagation described in Eq.~(\ref{eq:noiseSuppres}), it is not necessary to intentionally reduce the spatial resolution of the detected intensity images prior to applying the TIE by increasing \(L_{\mathrm{det}}\), in order to achieve an effective quantum noise reduction. At the same time, it is worth noting that the intrinsic suppression of high-frequency noise does not prevent the retrieval of high-resolution phase images. In fact, as it is clear from Eq. (\ref{tie}), in the direct-propagation problem, higher-frequency components of the phase object produce a stronger effect on the intensity. Thus, regions with rapid phase variations, i.e. high-frequency features, are usually reconstructed accurately despite the suppression introduced by solving the TIE \cite{Paganin_2004}.

Since low spatial frequencies dominate the overall noise contribution in the retrieved phase, the optimal correction factor in Eq.~(\ref{substitution}) for phase estimation can be evaluated analogously to the amplitude case, but in the limit of a large integration area. In this regime, $k^{(\text{TIE})}_{\text{opt}} = k^{(\tau)}_{\text{opt}}(\mathcal{D} \gg 1) \approx \eta_0$, thus losing its strong dependence on the resolution of the acquired intensity images. Further details, particularly regarding the validity conditions of this approximation, are provided in the M\&M section.

These results are shown in Fig.~\ref{HRQPI_Scheme}\textbf{c}. The four panels analyze the quantum advantage as a function of the phase resolution for four different defocusing distances \(dz\). The quantum advantage is quantified as the ratio between two Pearson correlation coefficients: one calculated using quantum-corrected phase images, \(\phi^{\text{qua}}(\mathbf{x})\) and the other using classical (shot-noise-limited) phase images, \(\phi^{\text{cla}}(\mathbf{x})\), each compared with a reference, \(\phi_r(\mathbf{x})\), obtained by averaging \(10^3\) frames to suppress residual shot-noise fluctuations. The classical and quantum averaged images, obtained by sending a total number of photons through the sample that is one thousand times larger than in the single-shot case, are indistinguishable for any practical purpose. From a numerical standpoint, either image can therefore be used equivalently as a noise-free reference. The Pearson correlation coefficient $\mathcal{C}$, a widely used metric to evaluate the similarity between two images, is defined as:
\begin{equation}
\mathcal{C} = \frac{\sum_{\mathbf{x}} \left(\phi_r(\mathbf{x})-\overline{\phi}_r\right)\left(\phi(\mathbf{x})-\overline{\phi}\right)}{\sqrt{\text{Var}[\phi_r]\,\text{Var}[\phi]}},
\end{equation}
where $\overline{\phi}$ and $\text{Var}[\phi]$ denote the spatial mean and variance of the phase image $\phi$, respectively.

In the figure, the blue data points represent the experimental quantum advantage obtained using the optimization factor $k^{(\text{TIE})}_{\text{opt}} = \eta_0$ in Eq.~(\ref{substitution}), with a single-channel overall detection efficiency estimated as $\eta_0 \approx 0.7$, while the red points indicate results obtained with $k^{(\tau)}_{\text{opt}}= \eta_0 \eta_c(\mathcal{D}) $, used for transmittance measurements and previously employed for phase imaging in Ref.~\cite{ortolano2023quantum}. The shaded red and blue areas represent results from Fourier-optics simulations \cite{voelz2011computational}, showing good agreement with the experimental data, whereas the green area indicates the gain in the quantum advantage achievable with the new reconstruction strategy.

For each defocusing distance \(dz\), the data are analyzed as a function of the phase spatial resolution. Since the phase resolution is directly determined by the spatial resolution of the intensity measurements, the latter is simply varied by increasing the effective integration area \(L_{\mathrm{det}}\) (i.e., \(\mathcal{D}\)). This is implemented by applying an averaging filter over neighborhoods ranging from 1 to 12 pixels.
The phase resolution is evaluated by fitting horizontal phase profiles, obtained from phase images averaged over $10^{3}$ frames, with an error function (Edge Spread Function, ESF). The spatial resolution is then defined as \(r_{\mathrm{phase}} = 2\sqrt{2\ln 2}\, w,\) where $w$ is the characteristic width parameter obtained from the ESF fit (see M\&M for better details). The averaging procedure over different frames is used to suppress shot-noise and isolate the intrinsic phase resolution. This choice allows us to disentangle the intrinsic spatial resolution of the optical system and of the TIE reconstruction from noise-induced effects.

For all values of $dz$, increasing \(\mathcal{D}\) degrades the phase spatial resolution. At the same time considering smaller  distances \(dz\), the finite-difference approximation of the axial derivative in Eq.~\eqref{tie} is more accurate, enabling a higher spatial resolution (see M\&M for details). In the opposite regime of large defocus distances, the finite-difference approximation deteriorates, resulting in a simultaneous reduction of both spatial resolution and quantitative accuracy of the phase estimation~\cite{ortolano2023quantum}.

We observe that using the correction factor $k^{(\tau)}_{\mathrm{opt}}(\mathcal{D})$, originally introduced for amplitude imaging, instead of \(k^{(\text{TIE})}_{\text{opt}}\), is counterproductive in the context of phase estimation. In particular, at the highest phase resolution, corresponding to small integration area \(\mathcal{D}\), the collection efficiency is $\eta_c(\mathcal{D}) \ll 1$, leading to $k^{(\tau)}_{\mathrm{opt}}(\mathcal{D}) \ll 1$. Consequently the quantum performance is comparable to the classical case, according to Eq. (\ref{substitution}). While this choice remains optimal for amplitude SSNQI, where it still yields the best noise reduction ~\cite{RuoBerchera_2020, ortolano2023quantum}, it is not suited for phase reconstruction.

In contrast, in NIQPI, high-spatial-frequency fluctuations are intrinsically suppressed by the solution of the TIE. This makes it more advantageous to employ the factor $k^{(\mathrm{TIE})}_{\mathrm{opt}} \approx \eta_0$, which preferentially optimizes noise suppression at low spatial frequencies, even when operating on high-resolution intensity images.

In particular, the quantum advantage is most pronounced at small defocusing distances $dz$, as indicated by the larger values of the correlation coefficient ratio, approaching 30\%. In this regime, phase-induced intensity variations are weaker, and shot-noise removal becomes essential to obtain high-quality images. Moreover, thanks to the better approximation of the finite-difference of derivative, for $dz = 0.0125$~mm and the minimum integration length \(L_{\text{det}}\) corresponding to the physical pixel size, the best spatial phase resolution of approximately $4\,\mu$m is achieved. Conversely, increasing the integration area leads to a progressive loss of resolution, with a maximum of approximately $15\,\mu$m, consistent with our previous results~\cite{ortolano2023quantum} (see M\&M for better details). Conversely, at larger defocusing distances \(dz\), the quantum advantage becomes less pronounced; however, in this regime, the finite-difference approximation of the axial derivative is less accurate, preventing a full exploitation of the optimal phase resolution achievable with the TIE. 


In Fig.~\ref{fig:ClassicalQuantumLine}, we present the classical and quantum reconstructions of both the phase and transmittance profiles. The reconstructions clearly demonstrate the effectiveness of the method in separating transmittance and phase information, as well as highlighting the quantum advantage. For the transmittance reconstruction a \(\mathcal{D} = 3.8\) obtained by averaging over 12 pixels was used in order to maximize the collection efficiency \( \eta_c(\mathcal{D}) \). This choice enabled the achievement of an NRF of 0.45 and a quantum advantage of 25\% in the standard deviation of the reconstructed amplitude. Note that, given the transmittance level of the mask,  $\tau=0.93$, and the large photon counts, the classical image already provides good retrieval, which limits the visual perception of the quantum advantage in the reconstructed image. However, this does not diminish the significant quantum advantage observed in the measured uncertainty over the transmittance. 

Conversely, the phase reconstruction are retrieved without any integration on the detector's area and with the new optimized correction factor $k^{(\mathrm{TIE})}_{\mathrm{opt}}$, revealing a visual quantum advantage.  The phase's values estimated for the smallest defocus distance \(dz\),  are listed in the caption. They show a good agreement with theoretical predictions and a reduced uncertainty for the quantum case. For higher \(dz\) the approximation in Eq.~(\ref{eq:approx}) is less accurate, and other technical issues in the alignment affect the phase estimation, as discusses in the Supplemental Document.

The overall spatial resolution at the micrometer scale and the ability of the method to operate effectively in the presence of both phase and absorption contrast demonstrate the readiness of this method for practical application, in particular in biological microscopy. In the following we proceed to apply the technique to biological samples to practically demonstrate this use.

\subsection*{Biological Sample}

\begin{figure*}[t]
	\centering
    \includegraphics[height=0.55\textwidth,width=\textwidth]{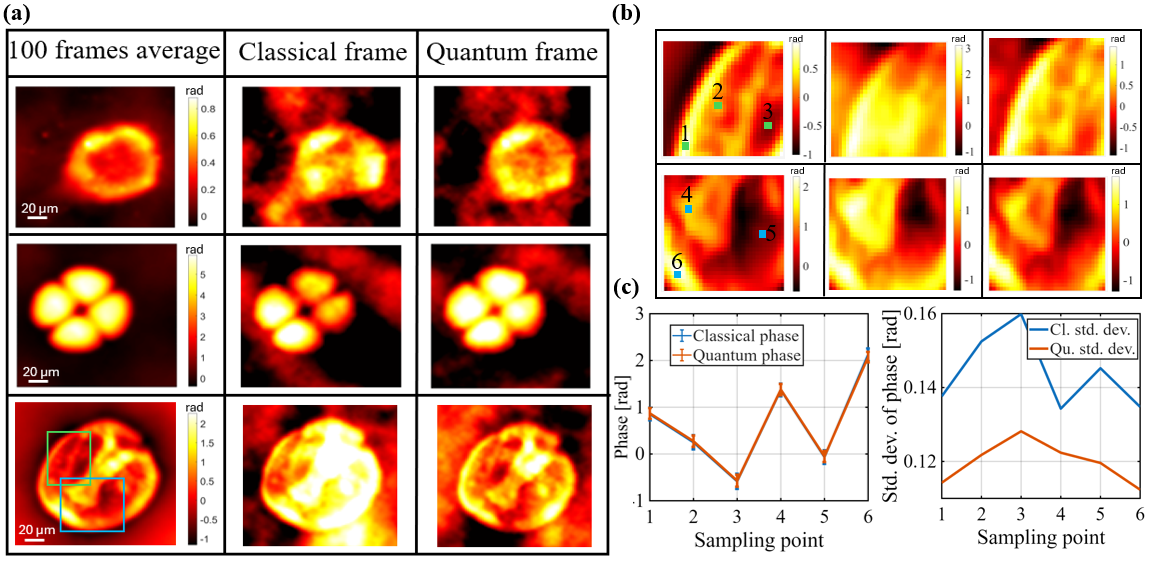} 
    \caption{\textit{Biological quantum phase imaging} (\textbf{a}) (Left column) Phase images obtained by averaging 100 frames, serving as a shot-noise-free reference. (Middle column) Classical single-shot phase images, affected by shot-noise. (Right column) Quantum-corrected single-shot phase images showing visibly improved quality and enhanced contrast compared to the classical case. Each row corresponds to a different biological sample, and the same color scale is used within each row. (\textbf{b}) Two areas, corresponding to the square boxes in last row of (a), are cropped. The color scales represent the entire range of the retrieved phase values. Green and blue dots are specific positions chosen for the quantitative analysis. (\textbf{c}) (Left) The average phase values corresponding to the six dots calculated over 100 frames. (Right) Standard deviation of the phase values in correspondence of the six dots. }
    \label{fig:biological}
\end{figure*}

We investigated unstained sea urchin ova, chosen for their natural transparency and intrinsic phase contrast. The samples were obtained from an industrial preparation with minimal staining, preserving the native optical properties of the cells.

Fig.~\ref{fig:biological}\textbf{a} presents wide-field phase reconstructions of ova at different developmental stages: from unfertilized eggs (top row), to fertilized eggs in early cleavage stages (middle row), to advanced embryonic forms (bottom row). These three biological conditions were imaged under different experimental settings. The defocus distances used in the TIE reconstruction were \( dz = 10\,\mu\text{m} \), \( 5\,\mu\text{m} \), and \( 5\,\mu\text{m} \), respectively, and the average number of detected photons per pixel was approximately 1000, 300, and 600 for the three samples.

The first column displays the reference phase image obtained by averaging 100 independent frames in order to eliminate any shot-noise. The second column shows the phase reconstruction from a single frame without any quantum correction. The effect of shot-noise in the phase reconstruction manifests as a low frequency cloud like patterns producing under- and over-exposed regions. The third column presents the quantum-enhanced phase image obtained by subtracting the shot-noise measured in the reference beam according to Eq.~(\ref{substitution})  with \( k^{(\text{TIE})}_{\text{opt}} =  \eta_0 \).

The quantum-corrected images demonstrate a visible improvement in contrast and noise suppression compared to the classical single-shot reconstructions, approaching the quality of the 100-frame averaged images. Two small portions of the last image, the ones inside the square boxes,  are zoomed in Fig~\ref{fig:biological}\textbf{b}. Even in these small details, and using a natural color scale, the improved fidelity of the quantum image is clearly appreciable: several details, lost passing from the 100 frames to the single classical frame, are restored in the quantum frame. Moreover, we have chosen six points in the image (green and blue dots) and reported the corresponding average phase values and their standard deviation in Fig.~\ref{fig:biological}\textbf{c}. While the average values are the same for the classical and quantum cases, the standard deviation is clearly lower for the quantum reconstructions, showing a quantum advantage in the range 20-25\% in the quantitative estimation of the  phase point-by-point.
Concerning the overall fidelity, which includes the recovered morphological information and is well captured by the Pearson correlation coefficient \(\mathcal{C}\), a quantum advantage between \(5\%\) and \(20\%\) is observed across the different biological samples. The variation in the measured advantage mainly reflects differences in sample absorption and thickness, which lead to different signal levels and thus to a varying impact of shot-noise. These results highlight both the robustness of the method in the presence of low signal levels and its capacity to provide high-fidelity phase reconstructions in biologically relevant conditions.

\section{Conclusion}

We have demonstrated a powerful sub-shot-noise technique in the domain of phase imaging, which does not suffer from any trade-off between spatial resolution and quantum gain in sensitivity. This enables the development of a microscope with diffraction-limited resolution and sub-shot-noise enhancement. In the microscopic system presented here, although we reach micrometer-scale resolution, the minimal achievable resolution does not yet match the diffraction limit of the imaging system. This discrepancy is due to technical limitations, the most relevant being the imperfect approximation of the derivative in Eq.~(\ref{eq:approx}) for the chosen value of \(dz\). Further improvements could be obtained by multi-distance approaches to TIE, which combine intensity measurements at both small and large defocus distances \cite{zuo2013transport}, by employing structured illumination techniques~\cite{zuo2017high} or simply by considering a smaller defocusing distance. This last approach would be possible just by a finer control of the translator stages. Additionally, in our setup, the diffraction limit is close to the physical pixel size of the detector, making it challenging to precisely evaluate the resolution near the theoretical limit. 

To demonstrate the applicability of our quantum-enhanced phase imaging method to real-world biological systems, we performed phase imaging of cells with details of a few micrometers, clearly showing less noise artifacts with respect to the classical measurement. We achieved a quantum advantage of up to 20 \% in terms of image fidelity, evaluate through the correlation coefficient, and up to 25\% in the quantitative phase uncertainty. Finally, our method is scanning free, requires only two shots for complete phase retrieval and do not need complicated interferometric setups. Filling the gap between proof-of-concept and real word applicability, this work represents an important step for the quantum sensing technology, where only a very limited number of protocols have reached an application level like ours. Several applications can already be envisaged whenever high illumination level can damage the sample or in fast imaging where the illumination level for the single frame should remain limited.

\section{Materials and Methods}
\subsection{Experimental layouts}
The experimental setup is based on a continuous-wave (CW) laser source (OBIS405 Coherent) operating at a wavelength $\lambda_p = 405\,\text{nm}$ with a beam width of $0.5\,\text{mm}$. Although the source is CW, it is triggered in a pulsed mode synchronized with the camera’s acquisition, with a maximum output power of 100 mW. The pump beam generates photon pairs via spontaneous parametric down-conversion (SPDC) in a 1.5~cm-long, type-II beta-barium borate (BBO) nonlinear crystal with transverse dimensions of 0.8~\(\times\)~0.8~cm\(^2\). The down-converted photons are filtered by an interferential filter centered at $800 \pm 20$~nm, selecting only photons near the degenerate wavelength $\lambda_d = 810\,\text{nm}$.

Spatial correlations are established at the crystal far-field (CFF) plane, formed by a lens with focal length $f = 1\,\text{cm}$. A second imaging lens with focal length $f = 1.6\,\text{cm}$ projects the far-field onto the CCD detection plane with a magnification factor of $M = 8$. The detector is a Pixis 400BR Excelon CCD camera (Princeton Instruments) with a $1024 \times 1024$ pixel array and $13\,\mu\text{m}$ pixel pitch. The pixel size on the object plane is approximately $1.5\,\mu\text{m}$. The camera operates in linear mode with high quantum efficiency ($>95\%$), low electronic noise, and full fill factor. 

\subsection{Characterization of Twin-Beams by NRF}

\begin{figure}[ht]
    \centering
    \includegraphics[width=\linewidth]{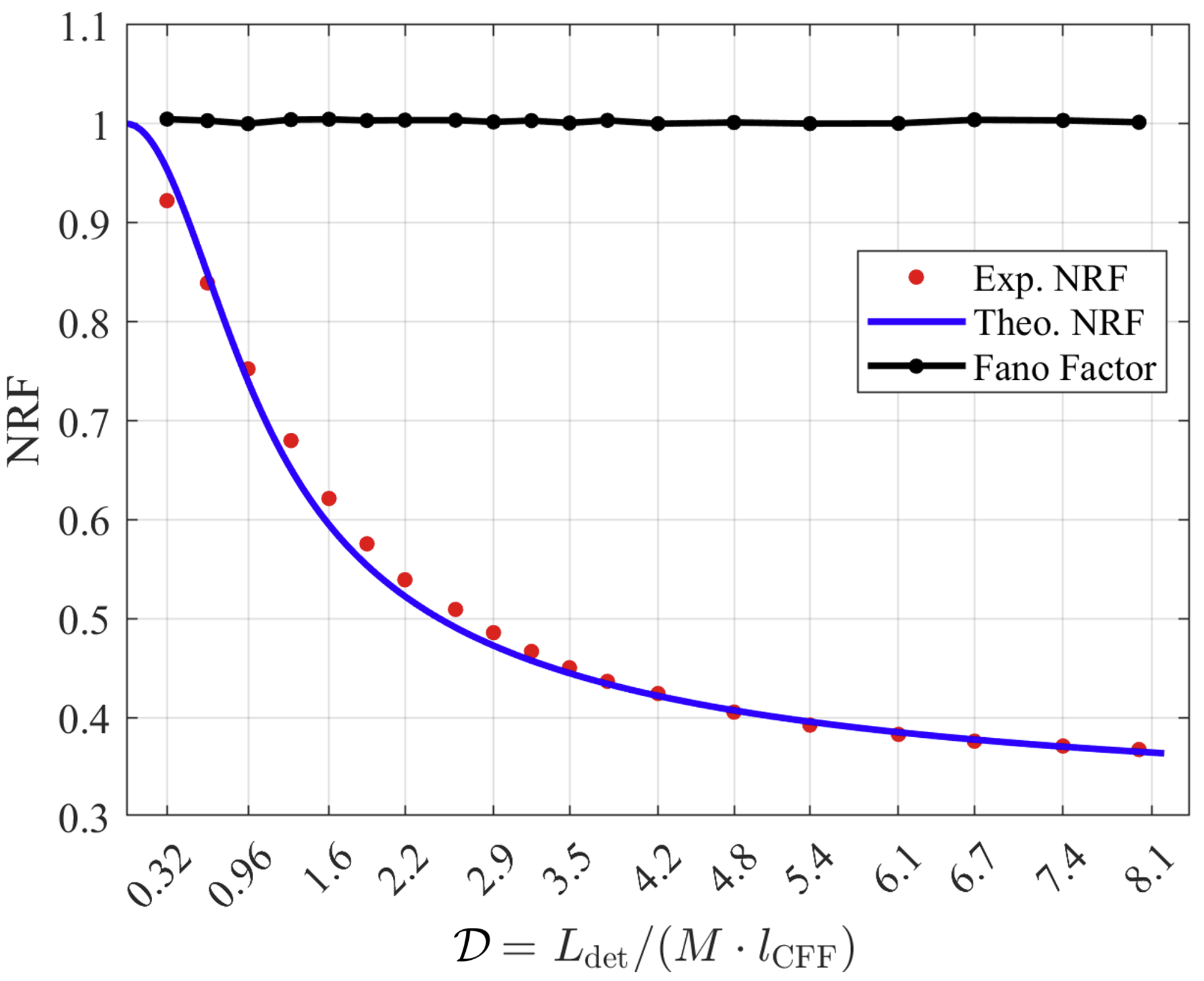}
    \caption{\textit{Experimental and theoretical behavior of NRF as a function of the parameter \(\mathcal{D}\).} Red points represent measured NRF values, the blue curve corresponds to theoretical predictions based on Eq.~\eqref{eq:NRF}, and black points indicate the Fano factor, confirming Poissonian photon statistics.}
    \label{fig:nrf_plot}
\end{figure}

We assume, as a good approximation, a  Gaussian spatial transverse correlation in the CFF, $\propto\,(2\pi\sigma^{2})^{-1/2}e^{-(\mathbf{x}_{i}+\mathbf{x}_{s})^{2}/2\sigma^{2}}$: if a signal photon is detected in the position $\mathbf{x}_{s}$ the twin idler photon will be detected according to that Gaussian probability around $\mathbf{x}_{i}=-\mathbf{x}_{s}$. The conditional detection efficiency \(\eta_c\)  depends on the pixel size (or the effective integration area) projected at the CFF, \(L_{\text{det}}/M\), where \(L_{\text{det}}\) is the linear size of the integration area and \(M\) is the magnification of the imaging system. It also depends on any misalignment, \(\Delta\), between the two pixels relative to their optimal positions. Specifically, \(\eta_c\) can be modeled as
\begin{equation}
\label{eq:etac}
\eta_c = \left(\frac{L_{\text{det}}}{M}\right)^{-2} 
\int_{\mathcal{A}} d\mathbf{x}_s \int_{\mathcal{A}} d\mathbf{x}_i \, 
\frac{1}{\sqrt{2\pi}\sigma} 
e^{-\frac{(\mathbf{x}_i + \mathbf{x}_s + \Delta)^2}{2\sigma^2}}
\end{equation}
where the integration area is \(\mathcal{A} =  L_{\text{det}}/{M} \times L_{\text{det}}/{M}\). The full width at half maximum (FWHM) of the correlation function is given by \(l_{\text{CFF}} = 2\sqrt{2\ln(2)}\, \sigma\).

By changing variables in Eq.~\eqref{eq:etac}, one finds that \(\eta_c\) depends only on two dimensionless parameters: the normalized detector linear size and the misalignment, both expressed in units of the correlation length. That is, 
\(\eta_c(\mathcal{D}, \epsilon)\), where \(\mathcal{D} = {L_{\text{det}}}/({M \cdot l_{\text{CFF}}})\) and \(\epsilon = \Delta / l_{\text{CFF}}\). For simplicity, we assume equal misalignment in both spatial directions, so that \(\Delta\) is treated as a scalar.
The conditional efficiency \(\eta_c(\mathcal{D},\epsilon)\) and the detection efficiency \(\eta_0\) were estimated through measurements of the Noise Reduction Factor (NRF), defined as \(\text{NRF} = \frac{\langle \delta^2 (N_s - N_i) \rangle}{\langle N_s + N_i \rangle}\). For SPDC light, the NRF can be approximated by:
\begin{equation}
\label{eq:NRF}
\text{NRF} \approx 1 - \eta_0 \eta_c(\mathcal{D}, \epsilon).
\end{equation}
Fig.~\ref{fig:nrf_plot} shows the experimentally measured NRF values (red dots) alongside the theoretical prediction (blue curve). The NRF was measured as a function of \(\mathcal{D}\), by varying \(L_{\text{det}}\). A fit of the experimental data to Eq.~\eqref{eq:NRF} was performed using two free parameters: the misalignment \(\epsilon\) and the single-channel detection efficiency \(\eta_0\). The fit yielded \(\epsilon = 0.2\) and \(\eta_0 = 0.7\), in good agreement with the experimental data. These parameters were used for the results shown in the main text.

The black points in the figure represent the Fano factor, \(F = \frac{\langle \delta^2 N \rangle}{\langle N \rangle}\), which remains close to one for all values of \(\mathcal{D}\), confirming that the photon statistics is in the Poissonian regime.

\subsection{TIE with phase and amplitude}

\begin{figure}[th]
    \centering
    \includegraphics[width=\linewidth]{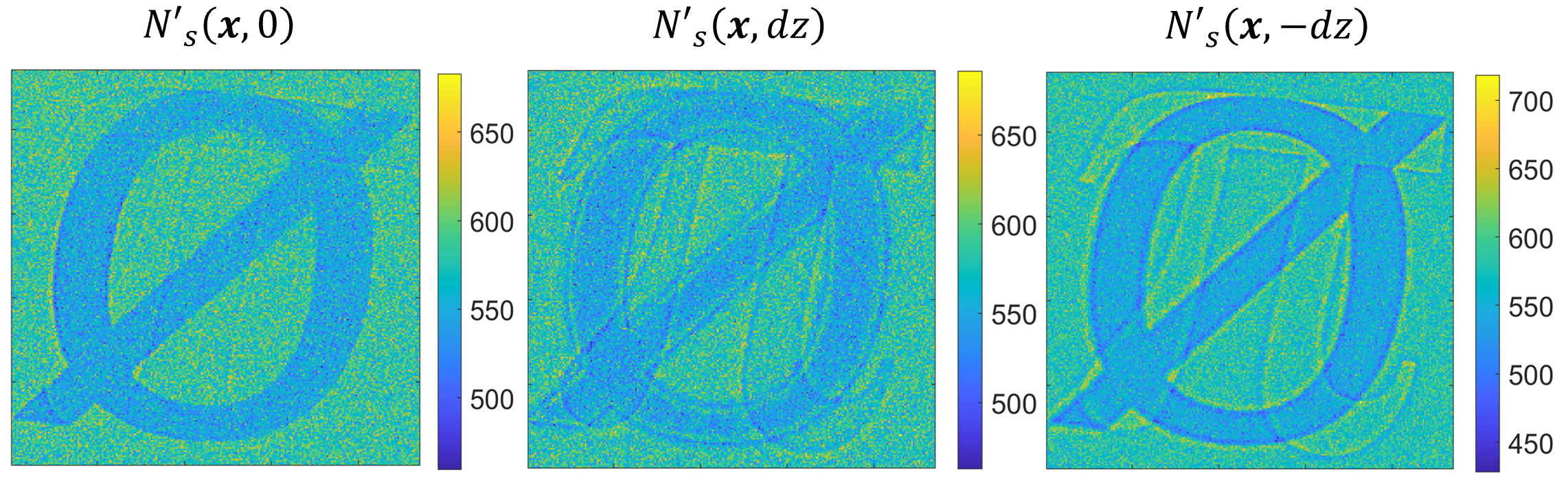}
    \caption{\textit{Photon counts with the object placed at three distances.} (Left panel) Photon counts recorded with the object positioned at the focal plane of the imaging system. In this case, only the transmittance mask (\(\varnothing\)) is visible. When the object is moved to defocused planes at distances \( \pm dz\) (middle and right panel), phase variations induce additional intensity modulations, resulting in an overlap of phase and amplitude effects. Nevertheless, the TIE allows decoupling of these contributions, enabling the quantitative and unique retrieval of the sample's phase profile.}
    \label{fig:Intensities}
\end{figure}

In Fig.~\ref{fig:Intensities}, the single-frame photon count images \(N'_s(\mathbf{x}, z)\) are shown for three different axial positions of the object relative to the CFF plane: \(z = 0\) (left), \(z = +dz\) (center), and \(z = -dz\) (right), with \(dz = 0.05\) mm. When the object is located at the nominal focus (\(z = 0\)), only the amplitude (non-uniform transmittance) features are visible in the intensity distribution. In contrast, at defocused positions (\(\pm dz\)), phase-induced propagation effects become evident, revealing the \(\pi\)-shaped phase structure superimposed on the amplitude modulation of the \(\varnothing\)-sign and its phase. Notably, the out-of-focus images exhibit overlapping features from both the amplitude and phase masks, underscoring the need for proper decoupling via the TIE to retrieve the pure phase profile.

A way to solve the TIE in Eq.~(\ref{tie}) with the measured photon counts is to adopt the so-called Teague’s assumption~\cite{Teague_1983}, introducing an auxiliary function \(\psi(\mathbf{x})\) so that
\begin{equation}\label{Teague}
I(\mathbf{x},0)\nabla \phi(\mathbf{x},0)= \nabla\psi(\mathbf{x}).
\end{equation}

According to Helmholtz’s theorem, the vector field \(I(\mathbf{x},0)\nabla \phi(\mathbf{x},0)\) can be decomposed into an irrotational component and a solenoidal component: \(I(\mathbf{x},0)\nabla \phi(\mathbf{x},0) = \nabla \psi + \nabla \times \mathbf{A}\), where \(\psi\) is a scalar potential and \(\mathbf{A}\) is a vector potential. It follows that Teague’s assumption holds if the transverse flux \(I(\mathbf{x},0)\nabla \phi(\mathbf{x},0)\) is irrotational. It has been shown that the irrotational approximation is valid if the in-focus intensity distribution is nearly uniform~\cite{Zuo_2020}. In this case, the phase discrepancy resulting from Teague’s auxiliary function is essentially negligible. However, when the measured sample exhibits low transmittance, the phase discrepancy may be relatively large and cannot be neglected. In our case, we consider a relatively high transmittance profile (see Fig.~\ref{HRQPI_Scheme}\textbf{b}), so we can safely use Teague’s assumption. The validity of the approximation is confirmed \textit{a posteriori} by the correct recovery of the phase values, as reported in the caption of Fig.~\ref{fig:ClassicalQuantumLine}.

Performing the substitution suggested in Eq.~(\ref{Teague}) into Eq.~(\ref{tie}) results in the following Poisson differential equation:
\begin{equation}\label{tie_psi}
-k\frac{\partial}{\partial z}I(\mathbf{x},z)=\nabla^{2}_{\mathbf{x}}\psi(\mathbf{x}).
\end{equation}

The solution to this equation is known, allowing the auxiliary function \(\psi(\mathbf{x})\) introduced by Teague to be determined. Subsequently, a second Poisson equation, directly obtained from Eq.~(\ref{Teague}), namely
\begin{equation}\label{tie_phy}
\nabla \cdot \left[I(\mathbf{x},0)^{-1} \nabla \psi(\mathbf{x})\right] = \nabla^{2}_{\mathbf{x}} \phi(\mathbf{x}),
\end{equation}
enables the reconstruction of a unique phase profile \(\phi(\mathbf{x})\), even in the presence of a weakly absorbing mask. This is done under Dirichlet boundary conditions, where the phase is set to zero at the borders of the reconstructed image.

\subsection{Phase resolution}
\subsubsection{Phase resolution enhancement}
\begin{figure}[th]
    \centering
    \includegraphics[width=\linewidth]{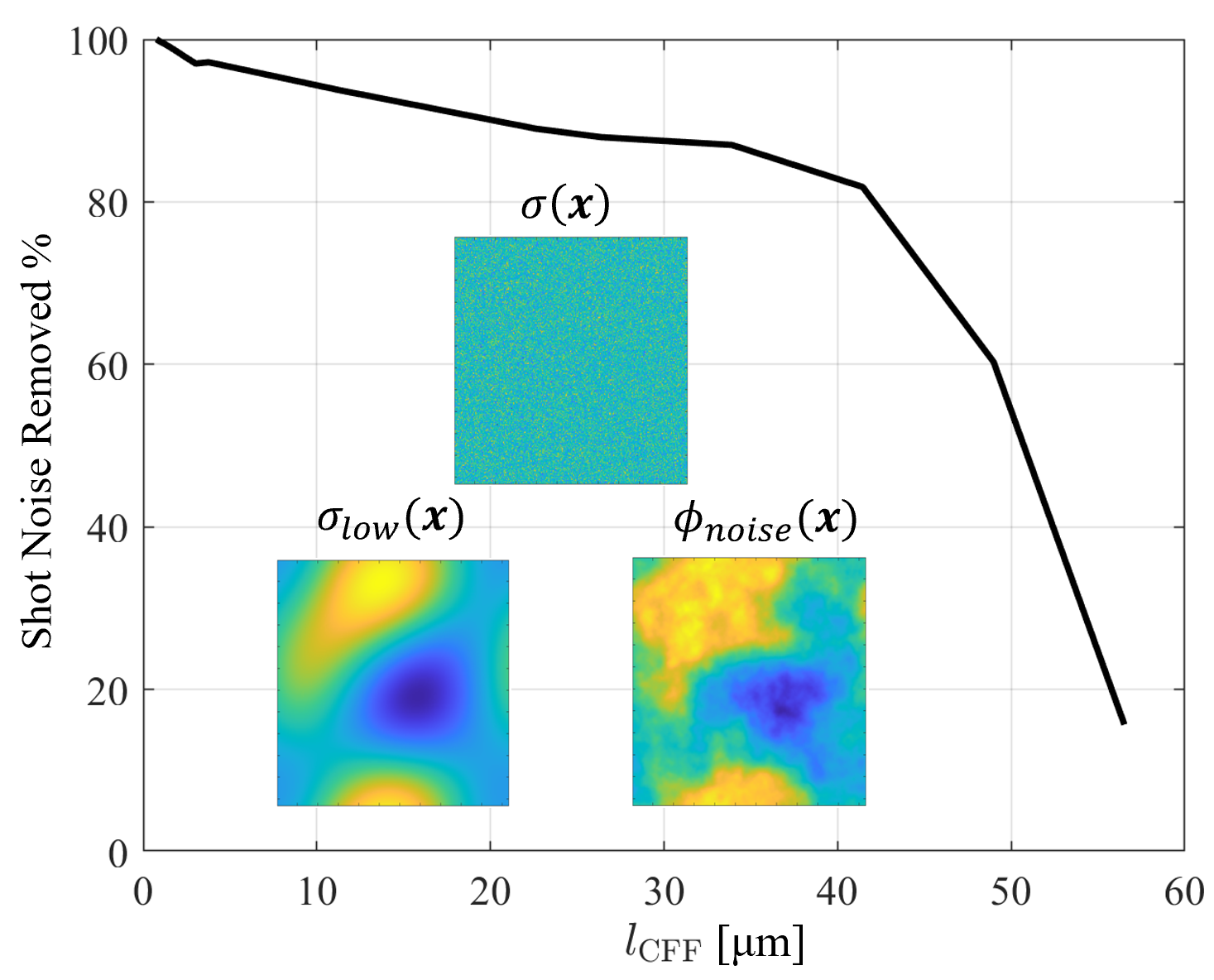}
    \caption{\textit{Visualization of noise suppression in phase retrieval with \(k^{(\text{TIE})}_{\text{opt}}\) estimator.} The plot shows the percentage of shot-noise removed as a function of correlation length \(l_{\text{CFF}}\). Insets show: (top) experimental Poissonian noise pattern \(\sigma(\mathbf{x})\), (bottom left) its low-frequency component \(\sigma_{\text{low}}(\mathbf{x})\), and (bottom right) the resulting phase noise \(\phi_{\text{noise}}(\mathbf{x})\) after TIE processing.}
    \label{fig:resolMatrices}
\end{figure}

A core aspect of this work is the demonstrated resolution enhancement. This improvement stems from the suppression of high-frequency noise via the solution of the TIE, as described in Eq.~(\ref{eq:noiseSuppres}). A visual explanation of this mechanism is provided in the inset images within the plot shown in Fig.~\ref{fig:resolMatrices}.

The central inset illustrates a typical noise pattern \(\sigma(\mathbf{x})\) (220~\(\times\)~220 \(pix^2\)), characterized by Poissonian shot-noise that affects the raw intensity measurements. The lower left panel shows the low-frequency component \(\sigma_{\text{low}}(\mathbf{x})\), obtained by spatial filtering to remove high-frequency content and reveal only slowly varying fluctuations. The right panel displays the corresponding phase noise \(\phi_{\text{noise}}(\mathbf{x})\), clearly showing that the dominant contribution to phase noise arises from low spatial frequencies. These components mainly affect broad regions of the retrieved phase image, approximately 100~\(\times\)~100 pixels in size.

The main plot in Fig.~\ref{fig:resolMatrices} reports simulation results evaluating the effectiveness of the TIE-based quantum method in suppressing phase noise as a function of the correlation length \(l_{\text{CFF}}\). For simplicity a uniform detection efficiency \(\eta_0=1\) is considered. In the simulation, starting from a noise distribution \(\sigma(\mathbf{x})\), a modified distribution \(\sigma'(\mathbf{x})\) is generated by redistributing photon counts across neighboring pixels using a 2D Gaussian kernel with a FWHM of \(l_{\text{CFF}}\), modeling the spatial spread of the correlations. From the corrected intensity \(\sigma(\mathbf{x}) - k^{(\text{TIE})}_{\text{opt}} \, \sigma'(\mathbf{x})\) the phase noise is retrieved solving the TIE and the resulting noise \(\phi_{\text{noise}}(\mathbf{x})\) is analyzed.

The curve shows the percentage of shot-noise removed as a function of \(l_{\text{CFF}}\). When the correlation length is small, \(l_{\text{CFF}} < 30~\mu\text{m}\), the TIE efficiently suppresses almost all of the noise. In particular, the experimentally estimated value \(l_{\text{CFF}} \approx 5~\mu\text{m}\) lies within this highly effective regime. However, as \(l_{\text{CFF}}\) increases beyond 40~\(\mu\text{m}\), the correlated noise begins to affect lower spatial frequencies, which are not suppressed by the TIE, and the quantum correction becomes less effective, leaving a substantial residual noise in the phase image.


\subsubsection{Phase resolution evaluation}

\begin{figure}[t]
	\centering
	\includegraphics[width=\linewidth]{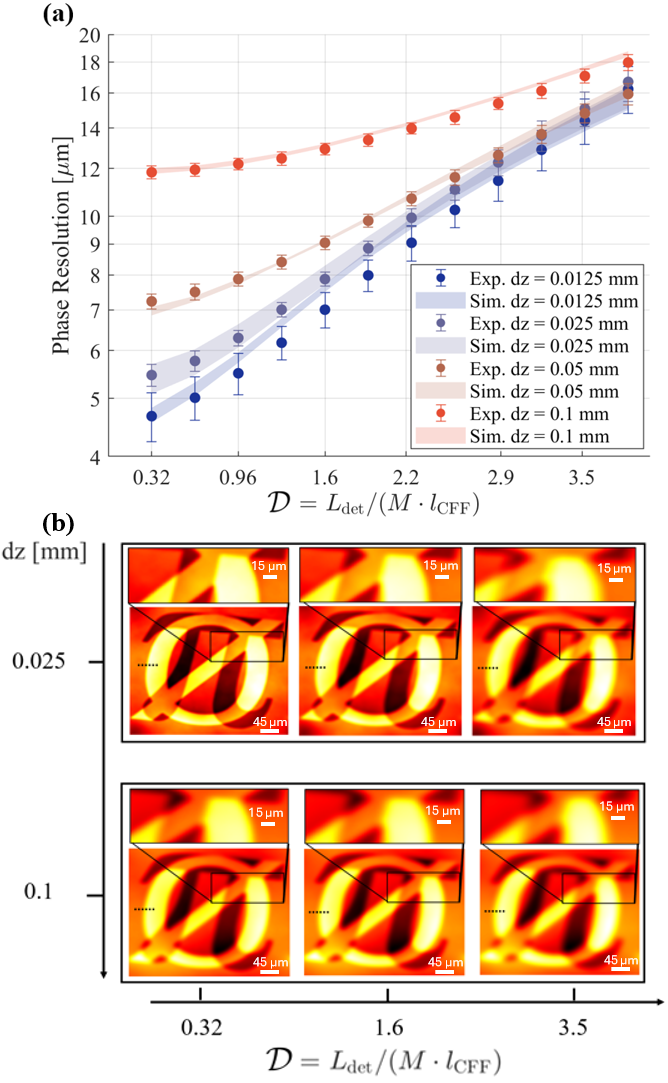} 
    \caption{\textit{Phase resolution evaluation with the optimized correction factor \(k^{(\text{TIE})}_{\text{opt}}\)} (\textbf{a}) Experimental and simulated phase resolution as a function of the resolution factor \(\mathcal{D}\) for different defocus distances \(dz\). The points correspond to experimental data for various \(dz\) values, while shaded areas represent theoretical predictions. As \(\mathcal{D}\) increases, the resolution worsens due to the larger effective pixel size, while smaller \(dz\) values allow better resolution. (\textbf{b}) Representative phase reconstructions for two different defocus distances (\(dz = 0.025\)~mm and \(dz = 0.1\)~mm) and three different \(\mathcal{D}\) values. Insets show magnified regions of the sample for a better qualitative evaluation of the phase resolution changes.}
    \label{fig:resolution}
\end{figure}

Fig.~\ref{fig:resolution} presents a quantitative evaluation of the spatial phase resolution achieved with the new \(k^{(\text{TIE})}_{\text{opt}}\)-based phase retrieval strategy, as a function of the resolution parameter \(\mathcal{D}\). The results, shown in Fig.~\ref{fig:resolution}\textbf{a}, are reported for different defocus distances \(dz\). Each data point represents the experimental resolution extracted from phase images averaged over \(10^3\) frames to eliminate residual shot-noise. The shaded area indicates the theoretical resolution limit achievable with the current optical system. Fig.~\ref{fig:resolution}\textbf{b} displays representative phase reconstructions, also averaged over \(10^3\) frames, for two defocus distances, $dz = 0.025\,\text{mm}$ and $dz = 0.1\,\text{mm}$, at three different values of \(\mathcal{D}\). Rectangular regions within each phase image are magnified to illustrate resolution variations more clearly.

As clearly shown by the curves in Fig.~\ref{fig:resolution}\textbf{a}, decreasing the defocus distance $dz$ improves the resolution by enabling a more accurate approximation of the derivative term in Eq.~\eqref{tie}, ultimately achieving a phase spatial resolution of approximately $4\,\mu$m for $dz=0.0125$ mm and for the minimal integration lenght \(L_{\text{det}}\) of the physical pixel, that correspond to \(\mathcal{D}=0.32\). The retrieved value is close to the theoretical resolution of the system, which results from the convolution of the effective pixel size at the object plane (1.7~\(\mu\)m) and the diffraction-limited resolution set by the numerical aperture (1.5~\(\mu\)m), yielding a combined resolution of approximately 3~\(\mu\)m. Conversely, increasing the integration area leads to a loss of resolution, where all the curves approach the same asymptotic behavior, with a spatial resolution of the phase images around \(18\,\mu\text{m}\), as achieved in our previous work~\cite{ortolano2023quantum}.

To evaluate the spatial resolution of our imaging system, we have analyzed the phase profiles extracted along a line perpendicular to the edge of the phase samples. These intensity profiles are fitted using an error function (Edge Spread Function, ESF) centered at position $x_0$:
\begin{equation}
    \text{ESF}(x) = \frac{a}{2}  \text{erf}\left(\frac{x - x_0}{\sqrt{2} \, w}\right) + b
\end{equation}
where $a$ and $b$ are fitting coefficients, and $w$ represents the characteristic width parameter of the transition region. Differentiating the ESF provides the Gaussian Line Spread Function (LSF):
\begin{equation}
    \text{LSF}(x) = \frac{d\,\text{ESF}(x)}{dx} = \frac{a}{w\sqrt{2 \, \pi}}\exp\left[-\frac{(x - x_0)^2}{2 \, w^2}\right]
\end{equation}
The spatial resolution of the system is defined by the FWHM of the LSF, given by:
\begin{equation}
    r_{\text{phase}} = 2\sqrt{2 \, \ln(2)}\,w
\end{equation}

To quantify the uncertainty, we first determine the 95\% confidence interval for the fitted parameter $w$, indicated as $[w_{\text{sub}}, w_{\text{sup}}]$. The standard error for $w$ is then approximated as $se_w = (w_{\text{sup}} - w_{\text{sub}})/3.92$ according to the $z$-test. Consequently, the standard error of the resolution $R$ is calculated as:
\begin{equation}
    se_r = \sqrt{2 \, \ln(2)}\,\frac{(w_{\text{sup}} - w_{\text{sub}})}{1.96}
\end{equation}

\section{Acknowledgments}
The authors are particularly grateful to Silvania F. Pereira for her helpful discussions and valuable suggestions for improving the paper.
This work has received funding from the
European Union’s through the following Projects: Horizon 2020 Research and Innovation Action under Grant Agreement Qu-Test (HORIZON-CL4-2022-QUANTUM-05-SGA);  Next Generation EU, Missione 4 Componente 1 CUP C63C22000830006-Bando a Cascata Spoke Fondazione Bruno Kessler-M4C2I.1.3.-NQSTI, and from the European Defence Fund (EDF) under grant agreement 101103417 EDF-2021-DIS-RDIS-ADEQUADE (Funded by the European Union). Views and opinions expressed are however those of the author(s) only and do not necessarily reflect those of the European Union or the European Commission. Neither the European Union nor the granting authority can be held responsible for them.

\section{Author contributions}

AP, GO and IRB devised the scheme of high resolution NIQPI. AP, GO and IRB developed the software for simulations and data analysis. The experiment and the data analysis have been carried out by AP with the help of IRB.  IRB and AP wrote the paper with the contribution of all the Authors. The non-biological sample has been designed, fabricated and characterized by SS. IRB and MG are equally last co-Authors. IRB supervised the project together with MG who is the director of the Research Sector in which the work has been developed.

\section{Conflict of interest}
The authors declare no conflicts of interest.

\bibliography{References/references} 

\end{document}